\documentclass[12pt]{article}
\usepackage{amssymb,epsfig}

\newcounter{multieqs}

\newcommand{\bq}{\begin{equation}}
\newcommand{\fq}{\end{equation}}
\newcommand{\bqr}{\begin{eqnarray}}
\newcommand{\fqr}{\end{eqnarray}}
\newcommand{\non}{\nonumber \\}
\newcommand{\noi}{\noindent}

\newcommand{\rf}[1]{(\ref{#1})}

\def\alp{\alpha}   \def\bet{\beta}    \def\gam{\gamma}
\def\del{\delta}   \def\eps{\epsilon} 
\def\zet{\zeta}        
       \def\lam{\lambda}
 \def\sig{\sigma}   
  \def\ome{\omega}

\def\cD{{\cal D}}  \def\cF{{\cal F}}
  
 \def\cK{{\cal K}} 
\def\cM{{\cal M}}

\def\pa{\partial}

\def\inv{^{-1}}

\def\pr{^{\prime}}

\newcommand{\tr}{\mbox{tr}}

\def\hlf{\frac{1}{2}}
\def\ove#1{\frac{1}{#1}}

\def\ZZ{\mathbb{Z}}

\def\bcomment#1{}
\begin{document}

\thispagestyle{empty}
\setcounter{page}{0}

%%%%%%%%%%%%%%%%%%%%%%%%%%

\begin{flushright}
\begin{tabular}{l}

CU-TP-{987} \\
NIKHEF-00-024 \\
YITP-00-43  \\
UPR-900-T \\

\end{tabular}
\end{flushright}

\bigskip

\begin{center}

{\Large \bf Dynamical Topology Change in M Theory}\\

\vspace{12mm}

{\large Brian R. Greene${}^{*}${}\footnote{E-mail:
greene@phys.columbia.edu},
Koenraad Schalm${}^{\sharp}${}\footnote{E-mail: kschalm@nikhef.nl}
and Gary Shiu${}^{\dagger,\diamondsuit}${}\footnote{E-mail:
shiu@insti.physics.sunysb.edu}}

\vskip 1cm

{${}^{*}$ {\em Departments of Physics and Mathematics,
Columbia University \\
New York, NY 10027}} \\[5mm]
{${}^{\sharp}$ {\em NIKHEF Theory Group\\
P.O. Box 41882 \\
Amsterdam 1009DB, The Netherlands}} \\[5mm]
{${}^{\dagger}$ {\em C.N. Yang Institute for Theoretical Physics,
State University of New York\\
Stony Brook, NY 11794}} \\ [5mm]
{${}^{\diamondsuit}$ {\em Department of Physics and Astronomy,
University
of Pennsylvania\\
Philadelphia, PA 19104}}{}\footnote{Address after September 1, 2000}

\vspace{5mm}

{\bf Abstract}
\end{center}

\noi
We study topology change in M theory compactifications on Calabi-Yau
three-folds in the presence of $G$ flux (the four form field strength).
In particular,
we discuss vacuum solutions in strongly coupled heterotic
  string theory
in which the
topology change is
{\em inevitable} within a single spacetime background. For rather
generic choices of
initial conditions, the field equations drive the K\"{a}hler moduli
outside the classical
moduli space of a Calabi-Yau manifold.
Consistency of the solution suggests that degenerate
flop curves --- just as wrapped  M theory fivebranes  --- carry magnetic charges
under the four form field strength.

\bigskip

\vfill

\begin{flushleft}
%\today
\end{flushleft}

\newpage
\setcounter{footnote}{0}

\section{Introduction}

Over the years, several works \cite{flop,conifold}
have established definitively
that there are  physically smooth processes in string theory
which result in a change in the topology of spacetime.
In these studies, as well as studies of topology change in M
theory \cite{Witten:1996qb},
{one considers} a one parameter family of vacuum
solutions --- a one parameter family of spacetimes --- {that}
passes from one Calabi-Yau manifold to another which is
topologically distinct. The referenced works succeeded in
showing that there is no obstruction to such topology change, but
no dynamics was ascribed to motion through the family.
In the present work, we study a variation on this
theme of topology change in which dynamics does
drive the evolution from one topology to another.
Specifically, (a) the topology
change occurs within
a single (not a family of) spacetime background and (b)
for generic choices of initial conditions, the dynamics
({\em i.e.}, the field equations) drive us through
a topology change.

To be concrete, we focus our studies on Calabi-Yau compactifications
of M theory to five dimensions in the presence of $G$ flux
(the four form field strength).
As discussed in \cite{strong,ovrut,gukov,warp}, the effective five
dimensional theory does not admit a flat space vacuum solution.
Rather, the spacetime metric is warped and the solution is
of the domain wall type with one of the five dimensions singled out as
the transverse direction.
In addition to the effective
five dimensional spacetime metric, the moduli of the Calabi-Yau
will generically
vary along the transverse direction.
In this paper, we show that there exist Calabi-Yau compactifications
in which
the field equations force
the K\"{a}hler moduli to
pass from one K\"{a}hler cone into an adjacent cone,
while the overall volume of the Calabi-Yau manifold remains large.
This implies that the Calabi-Yau manifold undergoes a flop transition
and continues on
to a topologically distinct Calabi-Yau manifold
as we move along the transverse dimension.

One may think of our work as being complementary to that
of \cite{blackhole1,blackhole2,lust} in which it was shown that in the
presence of
certain dyonic black holes, a Calabi-Yau with particular
moduli at spatial infinity can be driven by the attractor
  equations through a flop transition on the way to the black
hole's horizon. Here we find
vacuum solutions whose structure requires topology change.

One feature of these
topology changing solutions is that flop curves
appear to be magnetic sources for the $G$ flux.
This becomes apparent from the
Bianchi identity for the four-form $G$ in the context
of strongly coupled heterotic string theory (which we
shall henceforth refer to as Horava-Witten theory), i.e.
M theory on $S_1/\ZZ_2$.
In this set-up, the
 orbifold planes provide magnetic sources for the four form $G$
and
this requires a modification of the Bianchi
identity by a topological source term \cite{horava}.
As we will see, consistency of the topology changing solutions
 suggests
that the zero-size flop curve provides an additional
magnetic source for $G$.

While we will not pursue it in
this paper, the results of the present work may have
implications for the  ``brane world'' scenario
\cite{horava,ovrut,add,ST,BW,RS}. Strongly coupled heterotic
string theory
is
a rich context for the brane-world
scenario in which fairly realistic low energy models can be
constructed \cite{ovrut}.
The class of models of interest, however, is substantially enlarged,
as we allow the topology, and not just the metric,
of the Calabi-Yau to change from one end of the world to another.

This paper is organized as follows. In Section \ref{sec:review}, we
summarize the essential results of M theory and Horava-Witten theory
compactified
on
Calabi-Yau
three-folds in the presence of $G$ flux.
In Section \ref{sec:example},
we give an explicit example in which the solution to the field equations
inevitably drives us through a flop transition.
In Section \ref{sec:flop-transitions-m},
we discuss
general features of topology changing solutions in Horava-Witten theory,
and examine properties of $G$ flux and the Bianchi identity in such geometric
backgrounds.
We end with some discussion in
Section \ref{sec:discussions}.

\bigskip
After the presentation of these results by one of us \cite{Brian:talk},
M.
Wijnholt
and S. Zhukov notified us that they had also observed that the Bianchi
identity requires  modification in the presence of a flop transition.

\section{Kaluza-Klein Reduction of M theory on a Calabi-Yau three-fold}
\label{sec:review}

Let us briefly summarize the results of the Kaluza-Klein reduction of
eleven dimensional supergravity on a manifold
with boundaries, that is, strongly
coupled heterotic theory
on a Calabi-Yau three-fold $\cM$. (Many details can be found
in the papers
\cite{ovrut,ferrara,ceresole,gukov}; we differ in approach by using a
first order formalism for the field strengths, see also
\cite{Bergshoeff:2000zn}.
The appropriate Bianchi
identities are imposed as an additional field equation.)
We concentrate here only the
fields which are relevant to our subsequent discussion.
Since we seek solutions with non-trivial variation of
the K\"{a}hler moduli, we keep the
hypermultiplet scalar $V$ (the Calabi--Yau breathing mode) which
couples to the bulk potential terms, the
vector multiplet scalars $b^i$ (the K\"{a}hler moduli) as well as the
axionic field strengths $a_i$ and their dual potentials
$\tilde{\lambda}^i$. The
5-dimensional action can be consistently truncated
to this reduced field
content leading to \cite{ovrut}
\begin{eqnarray}
 S_5 &=& - \frac{L^6}{2\ell^9}\left[\int_{M_5}\sqrt{-g}\left(R+G_{ij}(b)
         \partial_M b^i\partial^M b^j+ \frac{1}{2}V^{-2}\partial_M V
         \partial^M V+\lambda (\cK - 1)
         \right)\right. \nonumber\\
      && \qquad\qquad
         \left. +\frac{1}{4} V^{-2} G^{ij}(b)a_i\wedge \star a_j
+d\tilde{\lam}^i
           \wedge a_i \right] -\sum_{n=0}^{N+1}
\alp^{(n)}_i\int_{M_4^{(n)}}(\tilde{\lam}^i
+\frac{b^i}{V}\sqrt{f^{\ast}g})~.
\label{S5_red}
\end{eqnarray}
Here ${\lambda}$ is a Lagrange multiplier,
{ the $a_i$ arise from the Kaluza-Klein reduction of
$G$ with respect to a basis of $H^4(\cM)$,} $G_{ij}$ is the metric on
the
K\"{a}hler moduli space (not the
5-dimensional metric which is denoted by $g_{MN}$; $f^{\ast}g$ is its
pullback to the three-brane worldvolume)
 and ${\cal K}$ its prepotential;
\begin{equation}
\label{eq:5}
G_{ij} =-\hlf \frac{\pa^2}{\pa b_i \pa b_j} \ln \cK ~,~~~~~
{\cal K} \equiv \ove{3!}d_{ijk} b^i b^j b^k ~,
\end{equation}
where $d_{ijk}$ are the intersection numbers of the (1,1) forms on
$\cM$.
The
gravitational coupling $\ell$ is that of the original
eleven-dimensional theory, and $L$ the length scale of the Calabi-Yau.

The boundary terms include the relevant contributions from $N$
five-branes at {\em a priori} arbitrary locations, with charges
$\alp_i^{(1)},\ldots, \alp_i^{(N)}$. In units of the five-brane
tension/charge
these correspond to the
multiplicity or equivalently the winding number of five-branes around
the various two-cycles. The first order action includes
the Bianchi identity, modified in the presence of five-branes, as
the field equation for the dual potential $\tilde{\lam}^i$
\begin{equation}
da_i \equiv J^{(5)}_i ~~\Rightarrow~~ \left\{\matrix{\displaystyle
\pa_{11}a_i
    =  \frac{2\ell^9}{L^6}
    \sum_{k=0}^{N+1}\alp^{(k)}_i\del(x^{11}-x^{11}_k)~, \cr
    \cr
    \pa_{\mu}a_i =0~,}\right.
\end{equation}
with solution \cite{strong,ovrut}
\begin{equation}
a_i = \frac{2\ell^9}{L^6}(\sum_{k} \alp_i^{(k)}\eps(x^{11}-x^{11}_k)
+c_i)~.
\end{equation}

The constants $\alp_i^{(0)}$, $\alp_i^{(N+1)}$ are the effective
M5-brane
charges carried by the end-of-the-universe 9-plane domain-walls. On
each end-of-the-universe plane, the effective five-brane charge has
two constituents, similar to the effective $Dp\pr,~(p\pr <p)$ charge
carried by $Dp$ branes in nontrivial backgrounds. One constituent is
the $E_8$ instanton number, which may be interpreted as M5-branes
immersed in the 9-brane \cite{strong,duff}; the second is the induced
five-brane charge due to the non-vanishing
curvature of the Calabi-Yau and proportional to $c_2(\cM)$
\cite{strong,ovrut}
\begin{equation}
\alp^{(k)}_i = {{T_5 L^{2}} \over {8 \pi^2}}\int_{D_i}  \left( \mbox{tr}
F^{(k)}\wedge
F^{(k)} - \hlf \mbox{tr} R \wedge R \right)~,~~~~~~~k=0,N+1.
\label{periods}
\end{equation}
Here $D_i$ is a 4-cycle of the Calabi-Yau,
and the elementary five-brane charge $T_5$ equals
\begin{eqnarray}
T_5 &=& \frac{2\pi}{(4\pi)^{2/3}\ell^6}~.
\end{eqnarray}
In the remainder we set all scales to unity.

The total five-brane charge must vanish on the orbifold interval
$S_1/\ZZ_2$;
\begin{equation}
\sum_{k=1}^N \alp_i^{(k)} +
\sum_{j=0,N+1} \int_{D_i}\mbox{tr}F^{(j)}\wedge F^{(j)}
=\int_{D_i}c_2~.
\end{equation}
This is the Kaluza-Klein reduction of the invariant
eleven-dimensional statement that the modified Bianchi identity
\cite{horava}
\begin{eqnarray}
dG&=&\left(\left[\tr F_{(1)}\wedge F_{(1)}-\hlf \tr R\wedge
R\right]\del(y)+\left[\tr
F_{(2)}\wedge F_{(2)}-\hlf \tr R\wedge R\right]\del(y-\pi R_{11})\right.
\non
&&\left.+\sum_in^i_5\del_{C^i}\del(y-y_i)\right)\wedge dy~,
\label{Bianchi1}
\end{eqnarray}
due to the presence of boundaries and five-branes is
integrable, {\em i.e.},
\begin{equation}
\int_{S_1/\ZZ_2 \times D_i} \hspace{-.2in} d G =
0~~~~ \forall ~~~~D_i \,\eps \,H_4(\cM) ~.
\end{equation}
In \rf{Bianchi1} $\{ C^i\}$ is a
basis of $H_2(\cM)$ dual to the basis $\{D_i\}$ of $H_4(\cM)$
and $\del_{C^i}$ is the four-form Poincar\'{e} dual to the 2-cycle
$C^i$ with delta-function support on $C^i$.  Preservation
of
supersymmetry allows only configurations of instantons and five-branes
(or anti-instantons and anti-branes).

If one chooses the standard embedding of the spin connection in the
first $E_8$ gauge group such that
\begin{equation}
\mbox{tr} F^{(1)} \wedge F^{(1)} = \mbox{tr} R \wedge R~,
\end{equation}
then $\alp_i^{(1)} = -\alp_i^{(2)}$, no additional five-branes are
needed and the effective five-brane charge on the Horava-Witten-plane
with unbroken gauge group at $y=\pi R_{11}$ is just the topologically
induced charge $\alp^{(2)}_i = - \hlf \int_{D_i}c_2$.

\bigskip

In this formulation the fields $a_i$ appear algebraically and may
be integrated out, yielding the $4$-dimensional domain-wall
action with electric $5$-form field-strengths $\cF^i =d\tilde{\lam}^i$;
\begin{eqnarray}
  \label{eq:44}
  S_5 &=& - \int_{M_5}\sqrt{-g}\left(R+G_{ij}
         \partial_M b^i\partial^M b^j+ \frac{1}{2}V^{-2}\partial_M V
         \partial^M V+\lambda (\cK - 1)\right)
          \nonumber\\
      && \qquad\qquad
         \left. -  V^2 G_{ij}(b)\cF^i\wedge \ast \cF^j \right]
         -\sum_{n=0}^{N+1} \alp^{(n)}_i\int_{M_4^{(n)}}(\tilde{\lam}^i
         +\frac{b^i}{V}\sqrt{f^{\ast}g})~.
\end{eqnarray}

The supersymmetric domain-wall or three-brane solution to the field
equations is given by \cite{ovrut}
\begin{eqnarray}\label{domainwallsoln}
ds_5^2&=&e^{2A}dx_4^2+e^{8A}dy^2 \non
V &=& e^{6A}~,\non
e^{3A}&=& \left( {1\over 3!} d_{ijk} f^i f^j f^k \right) ~, \non
b^i &=& e^{-A} f^i ~, \non
\cF^i_{11,\mu\nu\rho\sig} &=&\eps_{\mu\nu\rho\sig} e^{-10A}
\left(-\partial_{11}b^i+2b^i\partial_{11}A\right)~,
\end{eqnarray}
where the $f^i$'s are defined in terms of one-dimensional harmonic
functions
\begin{eqnarray}
\label{implicitsoln}
d_{ijk} f^j f^k = H_i ~, \quad \quad H_i &=& \sum_n \alp^{(n)}_i |y-y_n|
+
c_i \\
&=& \sum_{n=0}^{k} 2\alp_i^{(n)} y + k_i~,\quad\quad y_k<y<y_{k+1}
\end{eqnarray}
and the $k_i$ are arbitrary constants of integration.

The solution can be obtained from the supersymmetry
variations by searching for a
  global Killing spinor \cite{ovrut,gukov,Bergshoeff:2000zn}.
\begin{eqnarray}\label{killing}
\del \psi_{\mu}^A &=&\gam_{\mu}\left((\pa
A)\gam_{11}\eps^A-\frac{e^{4A}b^ia_i}{6V}(\tau_3)^A_{~B}\eps^B\right)
~,\non
\del\psi_{11}^A &=&\pa\eps^A-
\frac{e^{4A}b^ia_i}{12}(\tau_3)^A_{~B}\gam_{11}\eps^B ~, \non
\del\zet^A &=& (\pa \ln
V)\gam_{11}\eps^A-\frac{e^{4A}b^ia_i}{V}(\tau_3)^A_{~B}\eps^B ~,\non
\del \lam^{iA} &=& (\pa
b^i)\gam_{11}\eps^A+\frac{e^{4A}}{2V}\left(a^i-\frac{2}{3}b^ja_jb^i\right)(\tau_3)^A_{~B}\eps^B~.
\end{eqnarray}
The Killing spinor,
\begin{equation}
\eps^A = e^A(\del_B^A+\gam_{11}(\tau_3)^A_{~B})\eps_0^B~,
\end{equation}
is preserved, provided the following conditions are satisfied
  \begin{eqnarray}
    \label{eq:2}
        \pa A - \frac{e^{-2A}}{6}{b^ia_i}&=&0 ~,\non
        \pa b^i +\frac{e^{-2A}}{2}
\left( a^i -\frac{2}{3}~ (b\cdot a) b^i \right)
&=&0 ~.
  \end{eqnarray}
These are direct analogues of the attractor equations describing
spherically symmetric BPS black-holes in four- and five-dimensional
supergravities with eight supercharges
\cite{Strominger,Ferrara:1996dd,Ferrara:1996um}.
By contracting the second
equation in (\ref{eq:2}) with $\pa b^jG_{ji}(b)$, the third term
vanishes
whereas
 the first term yields the
length (in the moduli space) of the vector tangent to the path the
solution follows.
Defining a central charge $Z=b^ia_i$, and using the fact that the
$a_i=\alp^{(n)}_i\eps(y-y_n)$ are (piecewise) constant the resulting
equation exhibits
the monotonic flow of $Z$
\begin{equation}
  \label{eq:8}
  \pa Z
= - 2e^{2A}\pa b^i G_{ij}\pa b^j  \leq 0 ~.
\end{equation}
Alternatively one can define the dimensionful central charge
$\tilde{Z} = u^i \alp_i$ where $u^i
= e^{-2A} b^i$.
The attractor equation can be written in the form
\begin{equation}
  \label{eq:9}
  \pa \tilde{Z} = - \hlf\pa_i \tilde{Z} G^{ij}(u) \pa_j \tilde{Z}
\leq 0 ~,
\end{equation}
which relates the flow in the target space
with the flow in the moduli space.
Here,
\begin{equation}
\pa_i \tilde{Z} = \frac{\pa \tilde{Z}}{\pa u^i} ~.
\end{equation}

Notice that as in \cite{ovrut},
we are holding the hypermultiplets
(which control the complex structure) fixed, and
we shall continue to do so throughout this paper.
It would be interesting to relax this assumption as
one would need to do to study, for example,
conifold transitions.

\section{An Example}
\label{sec:example}

An essential aspect of (\ref{domainwallsoln}) is that the moduli of
the Calabi-Yau manifold will generically vary along the transverse
direction
$y$. This raises the question of whether one can find
solutions in which the K\"{a}hler moduli vary through a K\"{a}hler wall
in
the
moduli space; i.e. the Calabi-Yau manifold undergoes a flop
transition as we move along $y$.

Suppose we are able to find such a solution; this will have consequences
for
the analysis of the action
and the field equations in the previous section.
Specifically, notice that the equations (\ref{domainwallsoln}) and
(\ref{implicitsoln}) depend upon the second Chern numbers through
$\alp_i^{(0)}$ and $\alp_i^{(N+1)}$, and the intersection numbers
$d_{ijk}$ of
the Calabi-Yau. These numbers jump when a Calabi-Yau manifold
undergoes a flop. Hence, to exhibit an example where topology change
occurs,
we must search
for solutions that (a) pass through a wall
of the K\"{a}hler moduli space and (b) are a solution to the field
equations of the form (\ref{domainwallsoln}) for values of the
topological numbers $d_{ijk}$ and $c_2$
appropriate to the Calabi-Yau on {\em each} side of the wall.
Furthermore, in order to trust the supergravity approximation in which
we work, we must also ensure that the overall Calabi-Yau volume $V$
stays large.

In this section, through an explicit example involving some
fairly tedious algebra, we will show that
such topology changing solutions can indeed be found.
In this example we will choose the standard embedding
for the gauge bundles as this will facilitate
our discussion of the Bianchi identity and flop-curves
as sources of $G$-flux, in the next section.

A simple example of a pair of
Calabi-Yau manifolds connected by a flop transition
are the well studied
$(h^{1,1},h^{2,1})=(3,243)$ Calabi-Yau manifolds considered in
\cite{MorrisonVafa,Louis,CFKM,blackhole1,blackhole2}.
To investigate whether a flop can occur dynamically,
we will attempt to match the solutions to the field
equations for each of the Calabi-Yau manifolds across the singular
point. We
will find that, though the values of the fields can be matched
continuously, their first derivatives are discontinuous at the flop.
The jumps in the first derivatives of the K\"{a}hler moduli
imply that there
is an additional source of magnetic under the $G$ flux
charge where the solutions
are matched together (the flop point), and the K\"{a}hler
moduli are sensitive to this magnetic source. In the next section, we
will discuss the generality of this class of solutions.

Before we proceed, let us
collect the essential data of the Calabi-Yau and its flopped
cousin \cite{Louis}. Details
can be found in the aforementioned papers. In these works the
manifold we
denote with $\cM$ is known as
model III and $\widetilde{\cM}$ as model II. Both are elliptic
fibrations over Hirzebruch surfaces.

\subsubsection*{\it Calabi-Yau Manifold ${\cal M}$}
\label{sec:calabi-yau_org}

The intersection numbers are summarized in the classical prepotential
$\cK_{\cM}
=\ove{3!}d_{ijk}t^it^jt^k$. For $\cM$ the prepotential equals:
\begin{eqnarray}
\cK_{\cM}=\frac{1}{3!} \left( 8(t^1)^3+9(t^1)^2t^2+
                                   3t^1(t^2)^2+
                                   6(t^1)^2t^3+6
                                   t^1t^2t^3 \right)~.
\end{eqnarray}
The $t^i$ are the components of the K\"{a}hler form expanded in a
natural
basis (a basis in which the K\"{a}hler cone is defined by
$t^i>0$)
of 2-forms $\ome_i$ for $H^2(\cM,\ZZ)$,
\begin{equation}
J=t^i\ome_i~.
\end{equation}
Integrating $J$ over an arbitrary 2-cycle $C^i$ shows that the
corresponding $t^i$ measures its size.

The moduli $t^i$ are related to the dimensionless fields $b^i$
defined in the previous section by
$b^i=V^{-1/3}t^i$.
Therefore, the
dependence of $t^i$ on the transverse direction $y$ is
governed
by the one-dimensional harmonic functions $H_i(y)$.
The slopes of these harmonic functions are in turn determined by
the effective
five-brane charges.
For the end of the world
branes with standard
embedding, the slopes are proportional to the
periods of the second Chern class; in terms of the divisors
$D_i~\eps~ H_4(\cM)~$ dual to $\ome_i$, they are
\begin{equation}
c_2 (D_1) =  92 ~~,~~ c_2 (D_2) =  36 ~~,~~ c_2 (D_3) =  24  ~.
\end{equation}

\medskip

\subsubsection*{\it Flopped Calabi-Yau Manifold $\widetilde{\cal M}$}
\label{sec:calabi_yau_flop}

The intersection numbers
of this Calabi-Yau may be determined from its cousin $\cM$ with the
help of the relation (see, {\em e.g.}, \cite{flop})
\begin{equation}
\left( D_{i_1} \cap D_{i_2} \cap D_{i_3} \right)_{\widetilde{\cM}} =
\left( D_{i_1} \cap D_{i_2} \cap D_{i_3} \right)_{\cM}
- \sum_{\beta} \prod_k (D_{i_k} \cap C^{\beta}) ~.
\end{equation}
The $C^{\beta}$ represent the curves that are flopped and the $D_i$ on
$\widetilde{\cM}$ are the proper transforms of the $D_i$ on $\cM$.
In the present case there is only one such curve, and its intersection
with the $D_i$ is $D_i \cap C = \delta_{i3}$.
Hence, the only intersection number which
changes is $d_{333}$. It is shifted by $-1$.
Thus, if we choose the $D_i$ and their proper transforms
as bases on $\cM$ and $\widetilde{\cM}$ we find that
the prepotential of
$\widetilde{\cM}$ is equal to
\begin{equation}
\cK_{\widetilde{\cM}}=\cK_{\cM}-\frac{(t^3)^3}{6}~.
\end{equation}

The Chern coefficients $\tilde{c}_2 (D_i)$ of the flopped Calabi-Yau
$\widetilde{\cM}$ can
be calculated from
the Tian-Yau theorem in eq. (\ref{eq:99}) or using
the relation
\begin{equation}
\int_{D_i}c_2 + 2(D_i \cap D_i \cap D_i) = 12 \chi
(D_i) ~,
\end{equation}
and noting that the holomorphic Euler characteristic $\chi (D_i)$
is invariant under flops. Hence,
\begin{equation}
\tilde{c}_2 (D_1) = 92 ~~,~~ \tilde{c}_2 (D_2)= 36 ~~,~~ \tilde{c}_2
(D_3)= 26 ~.
 \end{equation}

\subsubsection*{\it Solutions to the field equations}

To find the explicit dependence of the moduli $f^i = V^{-1/6}t^i$ on the
transverse
direction we need to invert the nonlinear relations for $\cM$ and
$\widetilde{\cM}$
between $f^i$ and $H_i$.  {As mentioned}, we consider the case of
standard
embedding {\em i.e.}, $\mbox{tr} F\wedge F = \mbox{tr} R\wedge R$ at
$y=0$
and
$\mbox{tr} F\wedge F = 0$ at $y=\pi R_{11}$. In that
case the slopes $\alp_i$ of the harmonic functions $H_i$ are solely
proportional to periods of the second Chern class:
\begin{equation}
\alp_i = {1\over {16 \pi^2}}
 \int_{D_i} \tr R \wedge R =  \hlf\int_{CY} \ome_i \wedge c_2 (CY)~.
\end{equation}

First we choose a more convenient basis \cite{Louis, blackhole1},
\begin{eqnarray}
  \label{eq:bb}
  t^1 &=& U ~,\non
  t^2 &=& T-\hlf U-W ~,\non
  t^3 &=& W-U ~.
\end{eqnarray}
In this basis the respective
K\"{a}hler cones of $\cM$ and $\widetilde{\cM}$ are defined by the
regions
\begin{eqnarray}
  \label{eq:kc}
  \cM:&& W>U>0~,~T>\hlf U+W~, \non
\widetilde{\cM}:&& U>W>0~,~T> \frac{3}{2} U~.
\end{eqnarray}
and the area of the
flop curve
equals $W-U$.

The relation between the moduli and the harmonic
functions can now be straightforwardly inverted \cite{blackhole2}. For
$\cM$ one finds
\begin{eqnarray}
  \label{eq:inv}
 T&=& \hlf \frac{H_T}{U} ~,\non
U &=&\hlf\sqrt{(H_U+\hlf H_W)\pm \sqrt{(H_U+\hlf
  H_W)^2+2H_W^2-2H_T^2}} ~,\non
W&=& -\hlf \frac{H_W}{U}+\hlf U~,
\end{eqnarray}
with
\begin{eqnarray}
  \label{eq:harmm}
  {H}_T &=& 18|y|+{k}_T ~,\non
  {H}_U &=& 25|y|+{k}_U ~,\non
  {H}_W &=& -6|y|+{k}_W ~,
\end{eqnarray}
and for $\widetilde{\cM}$
\begin{eqnarray}
  \label{eq:inctM}
T&=& \sqrt{\hlf \tilde{H}_u\pm
  \hlf\sqrt{\tilde{H}_u^2-\frac{9}{4}\tilde{H}_T^2}} ~,\non
U&=& \frac{2}{3}\sqrt{\hlf \tilde{H}_u \mp \hlf \sqrt{
  \tilde{H}_u^2-\frac{9}{4}\tilde{H}_T^2}} ~,\non
W &=& \sqrt{-\tilde{H}_W}~,
\label{eq:10}
\end{eqnarray}
with
\begin{eqnarray}
  \label{eq:harmtm}
  \tilde{H}_T &=& 18|y|+\tilde{k}_T ~,\non
  \tilde{H}_U &=& 24|y|+\tilde{k}_U ~,\non
  \tilde{H}_W &=& -5|y|+\tilde{k}_W ~.
\label{eq:11}
\end{eqnarray}
Requiring that we are in the correct K\"{a}hler cone for
$\widetilde{\cM}$ selects the positive and negative sign
in the expressions for $T_{\widetilde{\cM}}$ and
$U_{\widetilde{\cM}}$ respectively.

\subsection{Matching the solutions}
\subsubsection*{\it Location of the flop}

If the field equations (whose behaviour within the framework on each of
the
Calabi-Yau's we have
already deduced) allow for
a topology changing transition, it must be possible to match the
solutions
across the point
where the flop-curve $t^3=W-U$ degenerates. As the relation between one
of
the moduli, $W$,
of the
Calabi-Yau $\widetilde{\cM}$ and the harmonic functions is rather
simple,
we will first
consider the flop solution from its point of view.

  From the data in \rf{eq:10}
and \rf{eq:11}
we see that at the flop point the following quantity must vanish
\cite{blackhole2}
\begin{equation}
9 \tilde{H}_W^2+4\tilde{H}_U\tilde{H}_W+\tilde{H}_T^2=0~.
\label{sec:Location-flop}
\end{equation}
Substituting the harmonic functions we find a quadratic equation for
$y$. This equation has a solution if the
discriminant is positive. The latter is a quadratic expression in the
three integration constants $\tilde{k}_i$ and vanishes at the two roots
\begin{equation}
\label{yflop}
25{\tilde{k}_U}=(45\pm\frac{5\sqrt{69}}{2})\tilde{k}_T+(42\pm9\sqrt{69})\tilde{k}_W
~.
\end{equation}
A brief inspection shows that large $|\tilde{k}_U|$ corresponds to a
positive
discriminant and  for a flop to occur we must therefore tune our
integration constants such that
\begin{equation}
25 \tilde{k}_U
>(45\pm\frac{5\sqrt{69}}{2})\tilde{k}_T+(42\pm9\sqrt{69})\tilde{k}_W
\end{equation}
or
\begin{equation}
25 \tilde{k}_U
<(45\pm\frac{5\sqrt{69}}{2})\tilde{k}_T+(42\pm9\sqrt{69})\tilde{k}_W ~.
\end{equation}

Regardless of the flop a necessary condition is that at the location
of the brane at $y=0$, we are using the correct Calabi-Yau data. One
condition, $2T>3U$, is guaranteed by the choice of solution in
\rf{eq:10}.
The
second one, $W>0$, is obeyed by requiring $\tilde{k}_W<0$. The last
constraint, $W<U$, yields a lower bound on $\tilde{k}_U$ in terms of
$\tilde{k}_T$ and
$\tilde{k}_W$
\begin{equation}
\label{inq}
\tilde{k}_U < - \frac{9\tilde{k}_W^2+\tilde{k}_T^2}{4\tilde{k}_W}~.
\end{equation}

Thirdly, by requiring that \rf{eq:10} has solutions,
we see that $\tilde{H}_U^2>9\tilde{H}_T^2/4$
or $2|\tilde{k}_U| >3|\tilde{k}_T|$. Fourthly, in order that
  the solution to (\ref{sec:Location-flop}) truly corresponds to a
  point where $U=W$, the lower bound
  \begin{equation}
    \label{eq:1}
    \tilde{k}_U> \frac{\tilde{k}_W}{2}
  \end{equation}
must be satisfied. Finally, in order that all divisors also have
positive area at $y=0$ one finds an additional constraint
$\tilde{k}_T>0$.

The parameter space that obeys all these conditions consists of
  two regions, and it is quite generic that a flop occurs for some
  choice of intitial conditions.
The two regions correspond to two possible flop scenarios:
(a) both roots, $y_{+}$ and
$y_{-}$, of
\rf{sec:Location-flop} are larger than zero or (b) both roots are less
than zero. {\em A priori}, there is the third possibility
that $y_{+}>0>y_{-}$. However, this
is not allowed because the quadratic \rf{sec:Location-flop} is convex
and
were we to find  $y_{+}>0>y_{-}$,
$W$ would be larger than $U$ at the location of the brane at
$y=0$. Hence we would be in the wrong K\"{a}hler cone.

Given the explicit values of the roots,
\begin{eqnarray}
  y_{\pm} &=&
\frac{1}{69}\left({10}\tilde{k}_U-18\tilde{k}_T-3\tilde{k}_W
\pm\sqrt{\cD} \right)~, \non
 \cD &=&
-612\tilde{k}_W^2-336\tilde{k}_U\tilde{k}_W+108\tilde{k}_T\tilde{k}_W
+100\tilde{k}_U^2-360\tilde{k}_U\tilde{k}_T+255\tilde{k}_T^2 ~,
\end{eqnarray}
the location of the minimum of \rf{sec:Location-flop} is
\bq
y_{min}
=\frac{1}{69}\left({10}\tilde{k}_U-18\tilde{k}_T-3\tilde{k}_W\right)~.
\fq
We will consider the region of parameter space such that
$y_{min}$ is positive.\footnote{As it turns out, the field equations do
not allow
topology changing solutions with standard embedding in scenario (b).}
It is easy to see that $y_{min}>0$
for large $\tilde{k}_U$ provided that the other
$\tilde{k}_i$ obey the above constraints
(for example, $k_U=100,k_T=30$ and $k_W=1$).
This region thus corresponds to scenario (a); the
Calabi-Yau manifold
at $y=0$ is model II with an $SU(3)$ gauge bundle and that at
$y=\pi R$ is model III with no gauge bundle.

\begin{figure}[t]
\begin{center}
{\epsfxsize=3in \epsfbox{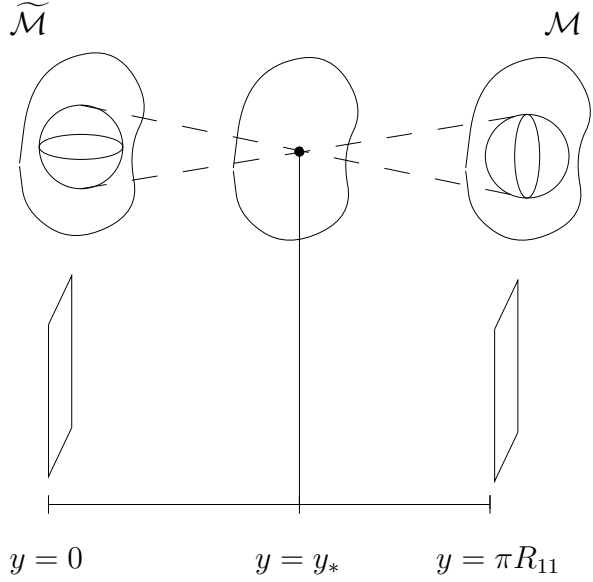}}
\end{center}
\vspace{-2.9in}\hspace{1.7in}{${\widetilde{\cM}}$}\hspace{2.6in}{${\cal
M}$}\vspace{2.6in}
\hspace*{1.7in}{$y=0$}\hspace{0.9in}{$y=y_{\ast}$}\hspace{0.5in}{$y=\pi
R_{11}$}\vspace{.1in}
\caption{Calabi-Yau configuration in heterotic M theory
which undergoes a flop.}
\end{figure}

\subsubsection*{\it Continuity}

We
demand that the solution for the 2-cycles $b^i = V^{-1/6}f^i$ can
be matched
across the flop at the flop location $y_{\ast}\equiv y_-$.
This implicitly requires that the harmonic
functions match  as well; $H_i(y_{\ast}) =\tilde{H}_i(y_{\ast})$.
As $c_2(D_T)=\tilde{c}_2(D_T)$
this implies that
$\tilde{H}_T(y) = H_T (y)$ for all $y$. Similarly
$H_U(y)+H_W(y)=\tilde{H}_U(y)+\tilde{H}_W(y)$. Combining this with the
final requirement
that
$H_U(y_{\ast})-H_W(y_{\ast})=\tilde{H}_U(y_{\ast})-\tilde{H}_W(y_{\ast})$
yields
\bqr
k_T&=&\tilde{k}_T ~, \non
k_U+k_W&=&\tilde{k}_U+\tilde{k}_W ~, \non
k_U-k_W &=&\tilde{k}_U-\tilde{k}_W- \hlf
y_{\ast}\left((c_2(U)-\tilde{c}_2(U))-(c_2(W)-\tilde{c}_2(W))\right) ~.
\fqr
In order that
$U_{\cM}(y_{\ast})-W_{\cM}(y_{\ast})$ indeed vanishes at the flop point
on Calabi-Yau $\cM$ with the above values of $k_i$,
we need to choose the negative sign in front of the
square root in \rf{eq:inv}.

It is now straightforward to plot the solutions on both side of the
flop for a particular choice of constants $k_i$ in region (a)
and establish
the occurence of a flop transition; see figure 2.

\begin{figure}[ht]
\hspace*{0.75in}
{{\epsfxsize=5in \epsfbox{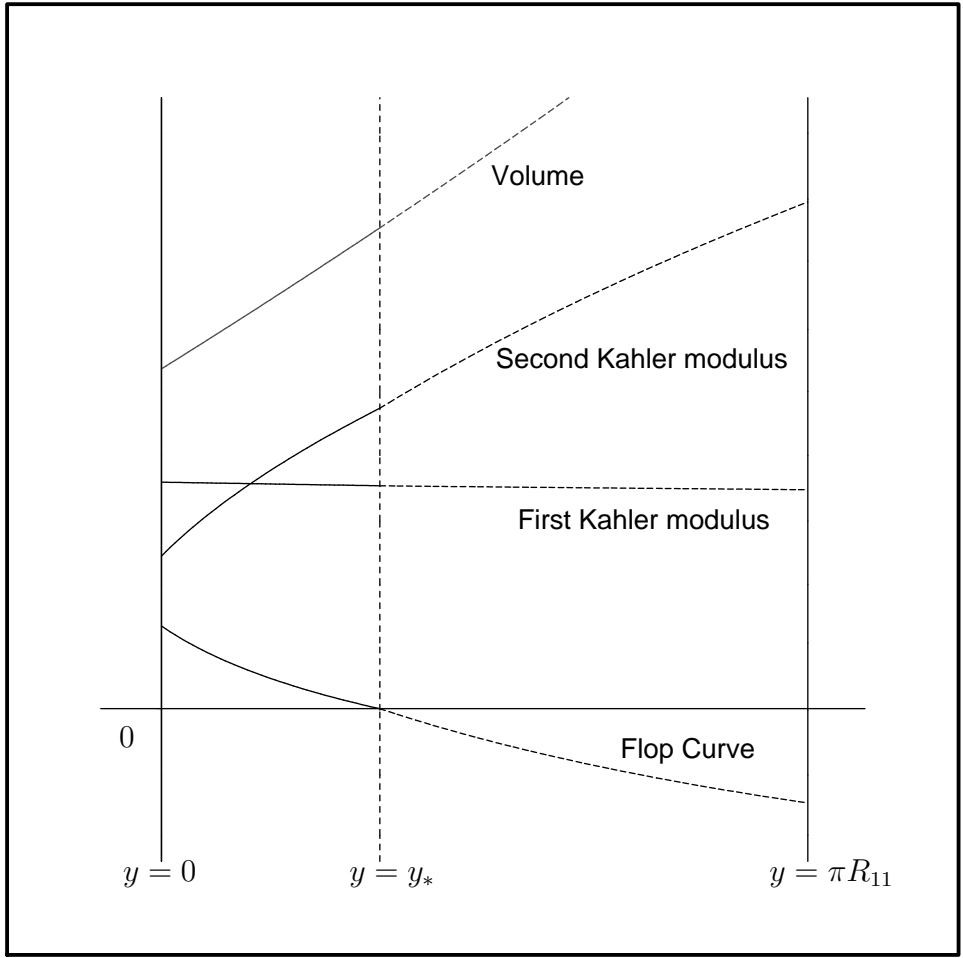}}}

\vspace*{-1.3in}
\hspace*{1.4in}{$0$}

\vspace*{0.5in}
\hspace*{1.42in}{$y=0$}\hspace{0.8in}{$y=y_{\ast}$}\hspace{1.75in}{$y=\pi
R_{11}$}
\vspace*{0.7in}
\caption{Profile of the areas of the two
cycles and the Calabi-Yau volume along $y$ in a basis appropriate for
Calabi-Yau $\widetilde{\cM}$. Solid lines belong to the
validity region $y<y_{\ast}$ of the Calabi-Yau $\widetilde{\cM}$. Dashed
lines belong to
the
Calabi-Yau $\cM$ and hold for $y>y_{\ast}$.}
\end{figure}

\subsubsection*{\it Discontinuity in the first derivatives}

We have seen that the function values of the moduli in the Calabi-Yau
manifold
${\cal M}$ can be continuously
 connected to that in
the flopped Calabi-Yau $\widetilde{\cal M}$ at the location
of the flop. We now check whether the first
derivatives are continuous.

In a universal, basis independent, notation the derivatives  of the
2-cycles $f^i$ are related to those
of the harmonic functions as
\bq
M_{ij} \pa f^j = \pa H_i.
\fq
Here $M$ is the symmetric matrix $M_{ij} = 2d_{ijk}f^j$.
This equation is nothing but the condition for the existence of a
global Killing spinor~(\ref{eq:2}) and therefore a component version
of the attractor equation~(\ref{eq:9}). Indeed $M_{ij}\pa f^j \sim
G_{ij} \pa b^j$.

In a natural orthogonal
basis such as the $t^i$ where exactly one of the $t^i$ shrinks to zero,
one
sees immediately that at the flop point
$M$ is continuous. The potential source of discontinuity, the jump in
the
intersection numbers,
always multiplies the very cycle which goes to zero. Recalling that
the derivative of the harmonic function is just the slope,
\bq
\pa f^j = (M\inv)^{ji}\alp_i~,
\fq
continuity
of the first derivative of the cycles rests solely on the continuity
of the slopes. Here,
knowing its relation to the moduli space metric,
 we assume that the matrix $M$ is invertible. The slopes, of
course, are the five-brane charges at each end of the universe. And
these
jump due to the flop transition:
\bqr
\pa f^j(y_{\ast})-\pa \tilde{f}^j(y_{\ast}) &=&
(M\inv(y_{\ast}))^{ji}\alp_i-(\tilde{M}(y_{\ast})
\inv)^{ji}\tilde{\alp}_i \non
&=& (M\inv(y_{\ast}))^{ji}(\alp_i-\tilde{\alp}_i)~.
\label{eq:3}
\fqr
Since $M$ is generically not (block-)diagonal, the first derivatives
of {\em all} cycles are discontinuous, as is evident in figure 2.
The reason for  the jump is the mismatch in $\int_{D_i} c_2 (R)$
on different sides of the flop.

Let us also briefly
check the behaviour of the volume $V$ of the Calabi-Yau,
\begin{equation}
V = \left( {1\over 3!} d_{ijk} f^i f^j f^k \right)^2 =
\frac{1}{36}(f^iH_i)^2~.
\end{equation}
It is obviously continuous for the same reason that $M$ is continuous.
As for
its derivative,
 it equals
\begin{equation}
\pa V = {\sqrt{V} \over 3} \left( \pa f^i H_i + f^i \pa H_i\right)~.
\end{equation}
Using the above result for $\pa f^i$ and the implicit relation
$f^i =\hlf M\inv(f(H))H$, this yields
\bq
6 \pa \sqrt{V} = \left( \pa H \cdot M\inv \cdot H + \hlf H \cdot M\inv
\cdot \pa H \right)~.
\fq
and one recognizes the first Killing spinor equation in
(\ref{eq:2}).
Again the expression is simplified at the flop; and one would guess
that the volume is discontinuous as well, as $H_i(y)$ and $M\inv(y)$
are continuous, but the slopes $\pa H_i(y)$ are not
\bq
\pa \sqrt{V}-\pa \sqrt{\tilde{V}} =
\ove{4}\left((\alp-\tilde{\alp})\cdot M\inv \cdot H\right)~.
\fq
However, using the implicit relation, $f^i =\hlf M\inv H$, once
more, this is seen to be
equal to
\bq
\pa \sqrt{V} - \pa \sqrt{\tilde{V}} =\hlf\left((\alp-\tilde{\alp})\cdot
f\right)~.
\fq
The only elements of the charge vector which jump are exactly those
parallel to the flopping curve $f^i_{flop}$, which shrinks to zero.
The potential source of discontinuity is therefore absent and
 $\pa V$ is in fact continuous. This also means that the second
derivative of $V$ has no delta-function singularity.

\subsection{Physical properties of the flop point}

\subsubsection*{\it Charge of the flop}

In comparing the
solutions of the two topologically distinct Calabi-Yau
manifolds at the flop, we find that the first derivatives of the
K\"{a}hler moduli,
$\partial_{11} b^i \sim \pa f^i$,
are discontinuous.

By virtue of our gluing together solutions whose {(piecewise)}
constant $G$ fluxes differ
in magnitude, there is a jump in $G$ flux on crossing the flop
point, {and such a jump will cause a discontinuity in the fields.}
As we will discuss in more detail in the next section, this is part of
why we will associate a magnetic $G$-charge with the flop point. More
generally,
though,
 for the $G$ flux to be globally defined,
we require that
\begin{equation}
\int d G = 0
\end{equation}
which implies that the sum of magnetic charges under the four-form
$G$ must vanish.
 If the topology of the Calabi-Yau does not change,
this condition is manifest for the standard embedding
configuration;
the geometrical five-brane charge induced from $tr R \wedge R$
is cancelled by
that of the instanton configuration.
However, as the topology of the Calabi-Yau on one ``end of the world''
changes, the global charge
conservation constraint is no longer satisfied.

Denoting the induced magnetic charges, $ \int_{D_i} G $, associated
with the
divisors $D_i$ of the Calabi-Yau manifolds by $(q_1,q_2,q_3)$, the brane
at $y=0$ reduced over the flopped Calabi-Yau $\widetilde{\cM}$
has induced geometric
charge $\alp^{y=0}_{geom} = -\hlf
\tilde{c}_2(D_1,D_2,D_3)=(-46,-18,-13)$
compensated by the
instantons with charge $\alp^{y=0}_{inst}\equiv
\tilde{c}_2(D_i)=(92,36,26)$. The brane at $y=\pi R_{11}$ only has
induced
geometric charges which equal $\alp^{y=\pi}_{geom}=-\hlf
c_2(D_i)=(-46,-18,-12)$. Adding all contributions we see that we
have an excess of $q_{total}=(0,0,1)$ units of charge.

The question is: what accounts for the missing (0,0,-1) units of
magnetic charge?
As indicated, a natural candidate is the flop curve which degenerates to zero
volume
at
the flop
point.
This situation is not completely unfamiliar.
There are examples where geometric singularities are
magnetic sources ({\em e.g.} orbifold singularities in
\cite{Morbifold}).
The singular object in the present setting is the degenerate
flop curve.
It is
therefore natural to conjecture that the flop
must in some way be accompanied by $(0,0,-1)$ units of charge to
  ensure global charge conservation.
\bcomment{
There is exactly one flopping curve at the location of the flop,
which can carry the missing $-1$ unit of magnetic charge.}
If this is so, the discontinuities in the solution to the field
equations
is just a consequence of the fact that
the flop carries magnetic charges, thereby inducing a jump in the
$G$-flux.

We will examine in more detail
the nature of the  charge at the flop point in the
next section.

\subsubsection*{\it Tension of the flop}

If the flop point is a charge source,
it is natural to inquire as to its possibly being a stress-energy
source as well.
The tension may be deduced from the curvature singularity.
For a metric of the form (\ref{domainwallsoln}), the non-vanishing
components of the Einstein tensor are
\begin{eqnarray}
G_{\mu \nu} &=& - 3 e^{-6A}
\left( {A^{\prime\prime} -2(A^{\prime})^2}
\right) \eta_{\mu \nu}~,
\non
G_{11,11} &=& -6 A^{\prime 2}~.
\end{eqnarray}
The delta function singularities of the Einstein tensor, indicating a
source
of tension, can only come from the term
$A^{\prime\prime}$, where $e^A = V^{1/6}$. In the previous subsection we
showed that the second derivative of $V$ is at most discontinuous.
Hence, the tension of the flop point is
zero. We note that this is in accord with the tension one would have
from a
five-brane wrapped around the flopping curve at the flop point.

\subsubsection*{\it Supersymmetry}

Finally, we should check whether the charge of the flop is consistent
with supersymmetry.
In order that the supersymmetries which are preserved all
have the same chirality,
{the inner products $f^i \alp_i^{(n)}$ for all charged
objects with charges $\alp_i^{(n)}$ must}
have the same sign (from the Killing spinor equations,
Eqs.~(\ref{killing}))
\begin{equation}
  \label{eq:12}
  f^i \alp_i^{(n)} \geq 0
\end{equation}
This inner product is equivalent to
$\int_{CY} J\wedge c_2(V)$ for vector bundles and $\int_{C^{(n)}} J$ for
five branes wrapping a
curve $C^{(n)}$. This gives the well
known condition that in order to preserve supersymmetry
the bundles must be holomorphic and stable,
and that five-branes must
wrap around holomorphic curves.
See also \cite{Bergshoeff:2000zn} for an in depth analysis
of the supersymmetry structure.

What about the proposed charge located at the flop curve?
As the volume of a flopped curve $\int_{C^{}} J$
vanishes at a flop point
it preserves the global Killing
spinor.

Note that since $f^i$ changes sign for the curve which flops,
we see from
eq.~(\ref{eq:12})
that  a supersymmetric
five-brane wrapping around the flopping
holomorphic curve on {either} side of the flop will have
{\em opposite} charge $\alp_i$.\footnote{Naively, one might think that
it is possible to introduce as many 5-branes wrapping around the
flopping curve
$C$ as possible, provided that we introduce an equal number of 5-branes
wrapping around the curve $\tilde{C}$ on the other side of the flop.
However, one has to check that the resulting
configuration still has a consistent topology changing solution
to the field equations.
\bcomment{since
the introduction of additional 5-branes changes
the solution to the field equations
(in particular the location of the flop).}}
This can be understood by noting that for a fixed set of divisors $D_j$,
if
the flopping holomorphic curve $C$ meets, say, $D_i$ transversely with
$C \cap D_i=+1$ in Calabi-Yau manifold $\cM$,
then the corresponding holomorphic curve $\tilde{C}$ on the flopped
Calabi-Yau
manifold $\widetilde{\cM}$ lies inside $D_i$,
with $\widetilde{C} \cap D_i=-1$.

\section{Topology Changing Solutions}
\label{sec:flop-transitions-m}

In the above example, we have seen that consistency of the solution
suggests that the zero-size flop curve is a source of magnetic
charge.
The effect of a magnetic charge at the
flop is to modify
the {Horava-Witten}   Bianchi identity  so that
$G$ is globally defined.
In this section, we 
discuss this idea in more detail.

Let us begin by considering 
the global charge
constraint anew. 
For simplicity, let us temporarily ignore the possibility
of
additional five-branes wrapped on other, non-degenerating, two-cycles in the
bulk.
In the eleven-dimensional theory on $S_1/\ZZ_2$,
the Bianchi identity for $G$ is modified by boundary sources
\cite{horava}.
We have,
\begin{eqnarray}
dG &=& \left( \left[\tr F_{(1)}\wedge F_{(1)}-\hlf \tr R_{(1)}\wedge
R_{(1)}
\right]
\del(y) \right.\non
&& +\left.\left[
 \tr F_{(2)}\wedge F_{(2)}-\hlf
\tr R_{(2)}\wedge R_{(2)}\right] \del(y-\pi R_{11})\right)\wedge dy  ~,
\end{eqnarray}
where we are now taking care to distinguish the curvatures of the
different Calabi-Yau spaces at each end of the interval. In the usual
case, where $\tr R_{(1)}\wedge R_{(1)}=\tr R_{(2)}\wedge R_{(2)}$, we
have
the familiar standard
embedding solution
\begin{equation}
\tr F_{(1)}\wedge F_{(1)}=\tr R\wedge R~~,~~\tr F_{(2)}\wedge F_{(2)}
=0~,
\end{equation}
to the global consistency constraint that
\begin{equation}
\int_{S_1/\ZZ_2 \times D} dG = \int_{S_1/\ZZ_2 \times
D}  dy ~\left(\hlf\tr
R\wedge R ~\del(y) - \hlf\tr R\wedge R ~\del (y-\pi R_{11}) \right) =0.
\end{equation}
But if $\tr R_{(1)}\wedge R_{(1)} \neq \tr R_{(2)}\wedge R_{(2)}$
(cohomologically) then the
mismatch implies that solely embedding either spin connection into the
gauge
group is no longer a solution.

A natural suggestion, then, is to seek out different holomorphic
stable bundles to  place at $y=0$ and $y=\pi R_{11}$ with different
second Chern classes, so as to find new consistent solutions to the
Bianchi identity. Indeed, one may be able to  find appropriate
bundles to restore anomaly freedom. However, for a number of
reasons, we propose a more general and universal solution.
 Namely, when $\tr R_{(1)}\wedge
R_{(1)} \neq
\tr R_{(2)}\wedge R_{(2)}$ because the Calabi-Yau has flopped somewhere
along $y$,
there is a new contribution to the Bianchi identity associated with the
collapsed flop curves.

To motivate our proposal, notice that given a potential
flop solution to
\rf{domainwallsoln},  {\em specifically} with standard embedding,
we can vary the location in $y$ at which the flop
occurs, by varying the integration constants $k_i$ --- as we have seen
explicitly in the previous section. Imagine now that we have such a
solution where a cycle shrinks to zero size
but we choose the $k_i$ such that the flop does not occur in
the physical range $0\leq y\leq \pi R_{11}$; rather, it happens
formally at $y=\pi R_{11} + \eps$. This is just the usual situation in
which the topologies of the Calabi-Yau manifolds at $y=0$ and $y=\pi
R_{11}$
are identical. Therefore, as expected, the choice of a standard
embedding of the spin connection into the gauge group is consistent.

Let's now adjust the $k_i$ so that the flop occurs at $\pi
R_{11}-\eps$, in the physical range so that the pure standard embedding,
with no other charge sources,
no longer provides a consistent solution. Where can the necessary other
charge
sources be? Four natural possibilities are: (1) Additional gauge
structure
at $y=0$, (2) New gauge structure at $y = \pi R_{11}$, (3) New wrapped
5-branes, or
(4) Magnetic charge associated with the geometrical singularity at
$y_*$.
By locality we do not expect to fix
the Bianchi identity by adjusting the gauge bundle at $y=0$, as that
lies at ``the other end of the universe'' from where the
flop occurs, making possibility (1) seem unlikely. Nor do we expect to
excite the new nontrivial gauge structure
by changing the geometry/topology of the Calabi-Yau
at $y = \pi R_{11}$ (where the initial gauge bundle
was chosen to be trivial).
In fact, as we will see in a moment, possibility (2) seems to be ruled
out
because the sign of the new magnetic source required to
fix the Bianchi identity conflicts with the requirement that
the new gauge bundle at $y  = \pi R_{11}$ is stable.
Possibility (3) immediately raises the question: where along
$y_{11}$ should the purported new 5-brane wrapping $C$ be located? From
the
discussion of the preceeding section, to contribute the correct charge,
it must be located at a point $y$ with $y \le y_*$. Locality, and
consistency
with the solutions we have constructed in section 3 in which
the flux jumps at the flop point, pick out $y = y_*$. We will
come back to this in a moment.

To understand possibility (4) we note that
since a flop causes the geometrical $G$-source
contribution $\tr R\wedge R$ to
change, it is natural to
suggest that a compensating magnetic source is provided by the
geometrical singularity at the location of the flop itself.
Strictly speaking, we are considering M theory compactified
on a $7$-manifold with boundaries, as the Calabi-Yau moduli
are varying over the $S^1/\ZZ_2$.
At the location of the flop,
the $7$-manifold is also singular --
with the singularity of the form
of a cone over $CP^3$ (see, {\em e.g.}, \cite{Liu})
\footnote{We thank Edward Witten and Chien-Hao Liu for discussions on this
point.}.
Geometrical singularities are known to carry
magnetic charges ({\em e.g.} orbifold singularities
in \cite{Morbifold}, and orientifolds).
The singular objects in the present setting are the flop curves when
they attain zero quantum volume.

To complete the proposal,
we need to specify the strength of the charge carried by the flop.
 A theorem of Tian and Yau, which we will discuss directly below,
suggests that, if a flop occurs at
$y=y_{\ast}$, the gravitational contribution to the Bianchi identity
is modified to
\begin{equation}
\label{eq:Bianc}
dG_{grav}= \left(-\hlf\tr R_{(1)}\wedge R_{(1)} ~\del(y)-\hlf\tr
R_{(2)}\wedge R_{(2)}~\del(y-\pi R_{11}) +\sum_{\bet}
\del_{C^\bet} ~\del(y-y_\ast) \right)\wedge dy~,
\end{equation}
where $\bet$ sums over all the holomorphic curves $\{C^{\bet}\}$ that
flop
at
$y=y_{\ast}$.

This additional gravitional contribution has a close
analogy with bulk five-brane sources wrapped
on nonsingular curves \cite{Morbifold}. The gauge bundle
contribution to $dG$ can be augmented by having five-branes wrapped on
two-cycles $C^i \subset \cM$ at locations $y=y_i$. This causes the
matter
part of the
Bianchi identity
to take the form \cite{strong} (see eq. (\ref{Bianchi1}))
\begin{equation}\label{eq:Bianchi5}
dG_{matter} = \left(\tr F_{(1)}\wedge F_{(1)}~\del(y) +\tr F_{(2)}\wedge
F_{(2)}
~\del(y-\pi R_{11}) +\sum_i\del_{C^i}~\del(y-y_i) \right)\wedge dy
\end{equation}
in the case of singly wrapped five-branes on holomorphic curves.

\bigskip

There is, however, an important subtlety in the above expression
(\ref{eq:Bianc}). In
the presence of a flop,
one has to
define precisely what one means by the curves $C^{\bet}$.
For with respect to a chosen basis of divisors $D_i$,
which (via invoking the proper transform map)
can be made universal on both sides of the flop, flopping curves differ
in
orientation from
one
side to the other. To be concrete,
suppose that the flopping curve $C^{\bet}$ and the
set of divisors $D_i$ meet transversely on a Calabi-Yau manifold ${\cal
M}$ with $C^{\bet} \cap D_i = 1$.
In the flopped Calabi-Yau manifold $\widetilde{\cal M}$, as we
mentioned earlier,
the flopping curve
$\widetilde{C}^{\bet}$
is no longer transverse to
$D_i$, but instead lies inside $D_i$,
with $\widetilde{C}^{\bet} \cap D_i=-1$.
Therefore, one has to specify in Eq. (\ref{eq:Bianc}) on which
Calabi-Yau
manifold the holomorphic curve $C^{\bet}$ is collapsing in
order that the geometrical data is complete.
In fact,
 this subtlety in assigning magnetic charges when a curve
collapses
is not intrinsic to the flop point, and is already present when one
considers a
five-brane wrapped around a collapsing curve.
We have discussed this subtlety
in the analysis of supersymmetry in the previous
section. In the same manner,
one has to specify on which Calabi-Yau manifold the
curve $C_i$ (that the fivebrane wraps) is collapsing.
The reason is that with
respect to
a fixed divisor $D_i$ (and its proper transform), the charge of the
five-brane differs by a
sign depending on whether one approaches the  flop point from one
Calabi-Yau or from its flopped cousin.

Let us examine more closely which holomorphic curve $C^{\bet}$ one
should
use in Eq. (\ref{eq:Bianc}).
Recall that our
topology changing solution can be obtained, starting from
a standard solution
in which the topology of the Calabi-Yau does not change
in the physical regime $0 \leq y \leq \pi R_{11}$
(which we call the ``parent'' solution).
Subsequently
by
varying the initial conditions of the two-cycles (the integration
constants $k_i$),
while keeping the gauge bundle fixed, we can bring the flop
point to $y=\pi R_{11}-\eps$.  This causes a
jump in $\tr R \wedge R$, which must be
compensated by the flop charge.
The question, then, is: What is the jump
in $\tr R \wedge R$ under a flop?
A theorem of Tian and
Yau \cite{TYau} states that starting from a Calabi-Yau manifold
$\widetilde{\cM}$ and a collection of holomorphic curves
$\{C^{\bet}\}$ on $\widetilde{\cM}$, the second Chern numbers of
$\widetilde{M}$ and its flopped cousin $M$ are related by
\begin{equation}
c_2({\cM})= c_2(\widetilde{\cM})+2\sum_{\bet}
\int_D[C^{\bet}]~,
\label{eq:99}
\end{equation}
with $D$ an arbitrary divisor and
$[C^{\bet}]\, \eps \, H^4(\widetilde{\cM})$, the Poincare dual of
$C^{\bet}$. We therefore see that
\begin{equation}
-\hlf \tr_{\cM} R_{(2)}\wedge R_{(2)}+\sum_{\bet}\del_{C_{(1)}^{\bet}} =
-\hlf
\tr_{\widetilde{\cM}}
 R_{(1)}\wedge R_{(1)}~.
\end{equation}
Hence, the $C^{\bet}$ in (\ref{eq:Bianc}) are holomorphic curves on the
Calabi-Yau manifold $\widetilde{\cM}$ of the ``parent'' solution.
The total $G$ charge, including that carried by the flop,
still equals
 $\tr R_{(1)}\wedge R_{(1)}$, which is exactly cancelled by the
standard embedding on the brane wrapping $\widetilde{\cM}$; the charge
conservation constraint
is satisfied.

Notice that the flop contribution to the Bianchi identity is
the same as that of a five-brane wrapping the flopping curve
on the Calabi-Yau of the ``parent'' solution
(which lies at $y=0$ in our setup).  However due to the fact that
supersymmetric five-branes wrapping the ``flopping'' curve
on the other ``flopped'' Calabi-Yau, must
carry the opposite charge,
the missing charge required to satisfy the Bianchi identity
{\em cannot} be
carried
by a five-brane
on the other side of the flop. Nevertheless, this means that
 the  magnetic source contributed by
a wrapped 5-brane over the appropriate degeneration of the flop curve
is identical to  the magnetic source of
the singularity at $y_*$ to which we are led
by the result of Tian and Yau. Hence,  at our level of analysis
 possibility (3) --- realized at $y_*$ --- and possibility
(4) are indistinguishable\footnote{An argument for possibility (3) is,
perhaps, that the above calculation indicates that the form of the local
singularity does not uniquely determine the sign of the charge it is
required to carry. Namely, as we have phrased the calculation, the
additional data of which Calabi-Yau is used in the "parent" theory is
needed. Equivalently, the sign of the charge depends upon the details of 
the background $G$-flux.}.

We can now also complete the promised argument that seems to rule
out possibility (2). The missing magnetic source for $dG$ in the
presence
of a single flopped curve is $\delta_C = -\delta_{C'}$ where $C$ lies on
the Calabi-Yau to the left of $y_*$ and $C'$ lies on the Calabi-Yau
to the right of $y_*$. The question then is whether there is a stable
holomorphic
vector bundle $V$ on the Calabi-Yau at $y = \pi R_{11}$ such that
\begin{equation}
c_2(V) = \hlf (c_2(T_2) - c_2(T_1)) = -\delta_{C'}~.
\end{equation}
But if there were such a bundle, then
\begin{equation}
\int c_2(V) \wedge J = -\int_{C'} J < 0
\end{equation}
violating the stability condition on $V$.
Hence, a gauge instanton at $y=\pi R_{11}$ alone cannot account either
for the
missing charge required to satisfy the Bianchi identity.

We note that  the
  inability of $G$-charge considerations
to distinguish between a five-brane of
the ``parent'' solution wrapping the zero-size
flop curve and a purely geometrical singularity raises the interesting
question whether flop transitions aside from new sources of
five-brane charge are also accompanied by the introduction of new
low-energy degrees of freedom \footnote{We note that preliminary
analysis indicates that we can deform the solutions of
section III so that the jump in the $G$-flux occurs to
the left of the flop point, with all quantitites then being
smooth across the flop point itself. This would naturally
be interpreted as the required $G$-charge being carried by
a 5-brane wrapped on the "would-be" flop curve, to the
left of the flop point. Further analysis of these
solutions would likely help clarify this issue.}.

We propose therefore, that in strongly coupled heterotic string
 theory, the general Bianchi identity for arbitrary
gauge bundles at $y=0$ and $y=\pi R_{11}$,
in a spacetime background in which the Calabi-Yau undergoes a flop
transition at $y=y_{\ast}$,
together with an arbitrary assortment of
5-branes wrapping two-cycles other than those involved
in the flop transition itself, is
\begin{eqnarray}
dG&=&\left(\left[\tr F_{(1)}\wedge F_{(1)}-\hlf \tr R_{(1)}\wedge
R_{(1)}\right]\del(y)
+\left[\tr F_{(2)}\wedge F_{(2)}-\hlf \tr
R_{(2)}\wedge R_{(2)}\right]\del(y-\pi
R_{11})\right.\non
&&\left.+\sum_in^i_5\del_{C^i}\del(y-y_i)
+\del(y-y_{\ast})\sum_{\bet}\del_{C^{\bet}}\right)
\wedge dy~.
\label{Bianchi}
\end{eqnarray}
with $C^i$ being holomorphic curves in the Calabi-Yau that the
corresponding five-brane
wraps, and $C^{\bet}$ being the flop curve
on the appropriate Calabi-Yau as described above. Of course,
the last two contributions can be grouped together, so long
as the contribution associated with any degenerating curves
is calculated according to the limiting procedure from the
correct side of the flop point.

\section{Conclusion}
\label{sec:discussions}

In this paper, we have studied topology changing solutions
in M theory compactifications
on Calabi-Yau three-folds with nonzero $G$ flux.
In the presence of $G$ flux, the field equations do not admit five
dimensional
Minkowski space as vacuum solution. The spacetime metric is warped,
with the solution of the domain-wall type so that one of the
five dimensions is singled out as the transverse direction.
In addition to the spacetime metric,
other moduli of the Calabi-Yau manifold also vary along the transverse
directions. We have studied an example in  strongly coupled heterotic
string theory --- the supersymmetric domain wall solution ---
in which the field equations
force us to go through a flop transition as we move
from one end of the world
to the other. Consistency of the
solution  suggests that
a flop curve --- like an ordinary (wrapped) M theory fivebrane --- carries
a single unit of magnetic charge under the $G$ flux.

We have focussed our attention on flop transitions.
As in \cite{ovrut} we have assumed that 
the hypermultiplets 
can be consistently decoupled in
the five-dimensional effective theory.
Especially in non-trivial geometric situations as presented here, it
would be interesting to revisit this issue of decoupling. Moreover, this
would allow the study  of more drastic
change of topology, such as conifold transitions, in which the
hypermultiplets necessarily play a role. As the number of moduli of the
Calabi-Yau changes across a
conifold transition, it is perhaps more appropriate to treat the
compact manifold as
a $G_2$ 7-manifold with boundaries which are
Calabi-Yau's. The low energy degrees of freedom in the effective
four-dimensional theory then correspond
to Ricci flat deformations of metric of the $7$-manifold.
We will leave this investigation for future work. From
a phenomenological point of view, it might be interesting
to look for more general solutions in
which the topology changes also with time.
This may have
implications for cosmology as well as other physics of the
brane world scenario.

\medskip

\noindent {\bf Acknowledgments}

\smallskip

We would like to thank Eric Bergshoeff, Jan de Boer, Mirjam Cveti\v{c},
Frederik Denef, Ron Donagi, Michael Douglas, Antonella Grassi, Sergei
Gukov,
Shamit Kachru,
Renata Kallosh, Chien-Hao Liu, David Morrison, Greg Moore, Burt Ovrut,
Jaemo Park,
Marco Serone, Stefan Vandoren and
Edward Witten
for useful discussions.
The research of B.R.G. was partially supported by the DOE grant
DE-FG02-92ER40699B.
The research of G.S. was partially supported by the NSF grant
PHY-97-22101
while he was at the C.N. Yang Institute for Theoretical Physics
at Stony Brook where this work was completed,
and was supported in part by
the DOE grant FG02-95ER40893 and the School of Arts and Sciences
Dean's fund
 at the University of Pennsylvania where this paper
was written.
K.S. is grateful for the extended
hospitality of the Spinoza Institute at Utrecht and the C.N. Yang
Institute for Theoretical Physics at Stony Brook.

\end{document}